\documentclass[conference]{IEEEtran}
\IEEEoverridecommandlockouts
\usepackage{cite}
\usepackage{amsmath,amssymb,amsfonts}
\usepackage{algorithmic}
\usepackage{graphicx}
\graphicspath{{figs/}}
\usepackage{tikz}
\usepackage{enumitem}
\usepackage{textcomp}
\def\BibTeX{{\rm B\kern-.05em{\sc i\kern-.025em b}\kern-.08em
    T\kern-.1667em\lower.7ex\hbox{E}\kern-.125emX}}

\usepackage{hyperref}
\newcommand{\note}[2]{{\color{red}#1: #2}}
\newcommand{\yousef}[1]{\note{yousef}{#1}}

\usepackage{xspace}

\begin{document}

\title{An Analysis of Home IoT Network Traffic and Behaviour}

\author{\IEEEauthorblockN{Yousef Amar}
\IEEEauthorblockA{Queen Mary University of London\\y.amar@qmul.ac.uk}
\and
\IEEEauthorblockN{Hamed Haddadi}
\IEEEauthorblockA{Imperial College London\\h.haddadi@imperial.ac.uk}
\and
\IEEEauthorblockN{Richard Mortier}
\IEEEauthorblockA{Cambridge University Computer Lab\\richard.mortier@cl.cam.ac.uk}
\and
\IEEEauthorblockN{Anthony Brown}
\IEEEauthorblockA{University of Nottingham\\anthony.brown@nottingham.ac.uk}
\and
\IEEEauthorblockN{James Colley}
\IEEEauthorblockA{University of Nottingham\\james.colley@nottingham.ac.uk}
\and
\IEEEauthorblockN{Andy Crabtree}
\IEEEauthorblockA{University of Nottingham\\andy.crabtree@nottingham.ac.uk}
}

\maketitle

\begin{abstract}
Internet-connected devices are increasingly present in our homes, and privacy breaches, data thefts, and security threats are becoming commonplace. In order to avoid these, we must first understand the behaviour of these devices.

In this work, we analyse network traces from a testbed of common IoT devices, and describe general methods for fingerprinting their behavior. We then use the information and insights derived from this data to assess where privacy and security risks manifest themselves, as well as how device behavior affects bandwidth. We demonstrate simple measures that circumvent attempts at securing devices and protecting privacy.

\end{abstract}


\section{Introduction}\label{s:intro}

There is an increasing presence of internet-connected devices in our homes, and yet we see an alarming rise in data thefts, security threats, and privacy breaches through these unregulated, uncertified, and often insecure devices.

With the recent discovery of the KRACK replay attack on devices using WPA2~\cite{vanhoef2017key}, the risks of unsecured communication within a network have once again surfaced to the public eye.

Similarly, with standards slow to catch up, manufacturers of networked devices have been making assumptions about the protocols they use, relying on undocumented behaviour. Standards to combat these issues do exist. For example, HTTP/2 requires encryption by default, blacklists insecure cipher suites, and is still faster than HTTP/1.1. Yet manufacturers are slow to adopt it and rely on legacy hardware and software.

TLS 1.3 uses ephemeral keys, making it more difficult for an attacker to decrypt traffic. This has seen pushback \cite{tls} as some heavily regulated enterprises (e.g.\ banking) rely on using static keys to decrypt and monitor internal traffic to meet both security and visibility requirements. This emergent use case could not have been predicted and solutions that will satisfy contexts with different concerns --- such as home vs enterprise --- are therefore slow to be established.

Along the same lines, some imminent standards, such as DNS Over HTTPS (DOH), seek to solve many privacy issues and prevent service providers from discriminating between different kinds of traffic. This would mitigate DNS-based internet filtering by ISPs/governments, as well as greatly enhance user privacy.

There are many potential measures that can be taken to mitigate the issues we discuss in this paper, however they have only very recently begun to be explored \cite{apnic}.

From an IoT perspective, in order to understand how to contain these devices, we need to first understand their behaviour in-the-wild and in realistic deployment scenarios. In order to secure the home of the future, it is critical to be able to automate the process of fingerprinting their network behaviour such that threats across arbitrary devices can be identified and mitigated.

We aim to understand the overall IO behaviour of devices in an average IoT-enabled house. Our intention is to understand the protocols used, data volumes, types, data rate and transmission frequency, etc. Our data should enable us to assess the potential costs of these devices in terms of bandwidth and their privacy and security threats.

Doing so will allow us to develop systems that maximise privacy and minimise leakage by default, and providing an interface for users to be able to understand, monitor, and control the flow of personal data in their homes~\cite{databox}.

We demonstrate simple yet significant measures to circumvent attempts at securing devices and protecting privacy, such as using passively observed API keys to hijack control of smart light bulbs, or tracking Apple devices despite MAC address randomization.

\newcommand*\circled[1]{\tikz[baseline=(char.base)]{\node[shape=circle,draw,inner sep=1pt,scale=0.8,fill=lightgray] (char) {#1};}}

\section{Set-up and Dataset}\label{s:data}

\begin{figure*}
	\centering
	\includegraphics[width=\linewidth]{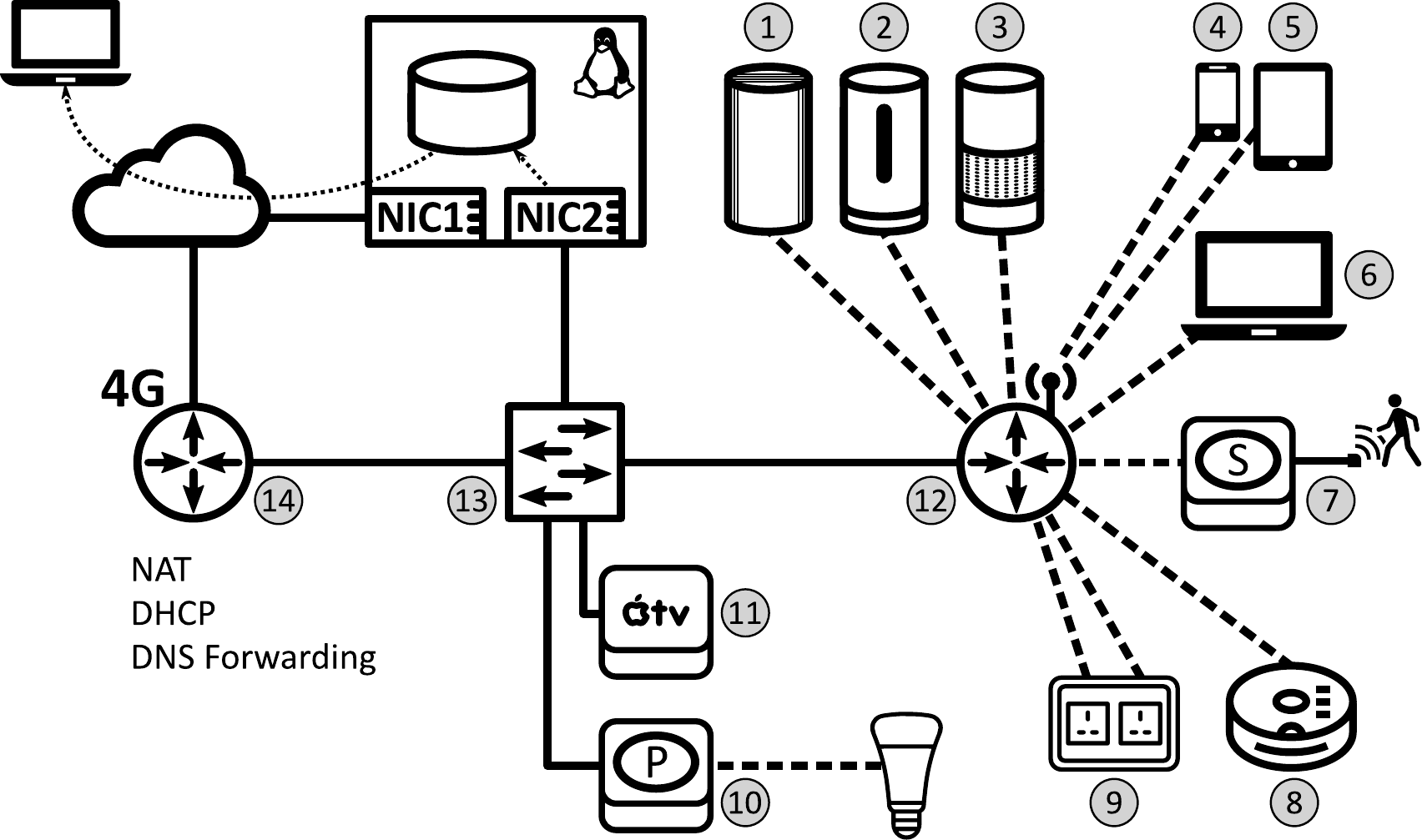}
	\caption{An Overview of the Home IoT Testbed}\label{f:testbed}
\end{figure*}

\subsection{Devices}

Figure~\ref{f:testbed} provides an overview of the IoT devices in our home testbed. The devices we used are as follows.

\begin{enumerate}[label=\protect\circled{\arabic*}]
	\item Foobot Air Quality Monitor
	\item Netatmo weather station and environmental sensors
	\item Amazon Echo
	\item Apple iPhone 5 running iOS 11
	\item Apple iPad pro 10.5" with cellular and WiFi connectivity
	\item Apple Macbook Pro 13" with touch bar
	\item Samsung SmartThings Hub with a presence sensor
	\item Neato Botvac vacuum cleaner
	\item 2x TP-Link smart plugs
	\item Philips Hue Bridge
	\item Apple TV
	\item Ubiquiti Access Point
	\item DLink Switch
	\item TP Link MR6400
\end{enumerate}

These include a variety of general-purpose home hubs (\circled{3}, \circled{7}), as well as device-specific hubs (\circled{10}, \circled{11}) that are one step away from the devices they actually control, such as several Hue LED bulbs.

We also have a number of consumer electronics connected (\circled{4}, \circled{5}, \circled{6}), while our two smart plugs (\circled{9}) can accommodate any additional offline devices. Finally, we have IoT devices that connect directly to the router (\circled{1}, \circled{8}, \circled{2}).

\subsection{Set-up}

This section describes how we set up our home IoT testbed. Our measurements and analysis is conducted on network traffic while all devices are idle. We connect an L2 switch to a 4G router running NAT, DHCP, and DNS forwarding to Google's DNS. An access point is connected to this switch to which all but two of the IoT devices are connected. These two have a wired connections directly to the from two of our IoT devices.

We capture all traffic by mirroring all ports to one connected to a Linux box with two NICs. Traffic is captured via one NIC with TCP/UDP disabled through tcpdump and stored on disk. Traces are then retrieved separately through SSH via a second NIC connected to the internet. All data must pass through the switch (both internal and external) and thus all packets are mirrored on the switch port to the Linux box and therefore all packets are captured.

\subsection{Data and Analysis}

We continuously captured packets for 22 days before performing our first analysis of this data. We wrote a set of scripts to perform our analysis, and are making these scripts publicly available \cite{github}. To analyze network behavior on a per-device basis, we split the combined trace by MAC address. We also used DNS and DHCP logs to help find hostnames that correspond to MAC addresses by looking at mappings of IP to MAC address and IP to hostname over time. This is useful especially for devices that randomize their MAC addresses, such as the iPad.

For all other statistics, we fed the traces through the Bro Network Security Monitor \cite{bro} and ran custom as well as existing scripts on Bro logs.

\section{Observations}
\label{s:results}

\subsection{Device Setup and Interaction}

The devices were set up using the manufacturers apps and set up with Apple HomeKit where possible. If a device could work with Alexa then it was configured to do so (Foobot, Neato Botvac, Philips Hue, and TP-Link plugs).

When setting up devices, it is not uncommon to go through a Bluetooth-like pairing flow (e.g.\ Apple TV). Some devices require WPS-like physical interaction through a button press (e.g.\ Philips Hue Bridge), however we show that this can be circumvented. Previous work has shown that the secure setup, pairing, and configuration of devices can be done securely in way that does not impact user experience negatively, so arguments for compromising security in favor of usability are weak \cite{brown2013multinet}.

Some devices, such as the Hue Bridge or the Neato Botvac communicate via plain HTTP with other devices and the outside world. This interaction can be monitored, device behaviour inferred, and in the case of the Hue Bridge, API keys can be extracted and the device hijacked.

\subsection{Learning from Traffic}

\subsubsection{Identifying Devices}

While for most devices, the first three MAC address bytes are enough to identify a vendor, and thus make a reasonable assumption as to what the device could be, there are many other methods that we explored in part during this experiment that can be used in conjunction.

For example, the Foobot's MAC address points to ``Shanghai High-Flying Electronics Technology Co., Ltd'' a WiFi module manufacturer. This tells us very little about what the device could be, but if we look at the DNS requests it makes, most of which are A record queries for api.foobot.io, we can build a behavioural profile for this device, and use it to identify others like it.

Similarly, consumer Apple devices can randomise MAC addresses to curb tracking, yet there are ways to track these devices regardless or even reveal their true MAC address \cite{vanhoef2016mac}. We were able to track an iPad connecting to our network by combining DHCP and DNS logs to ascribe the iPad's local hostname to all eight of the MAC addresses it used.

\subsubsection{Monitoring Devices}

While encrypting web traffic protects a user's data to some extent, there are still other risks that are not always obvious. For example, using a proxy will not always curb DNS leaks unless configured for remote DNS. A lot of information can be gleaned from DNS traffic, though fortunately efforts to protect this information (mainly from ISPs) are increasing. Recent commits to Android indicate that Google plans adding DNS over TLS to the ubiquitous operating system \cite{xda}.

The main issue however is that some devices still use plain HTTP for some or all of their communication (figure~\ref{f:traffic}). Any device on the network can passively observe metadata and control requests coming from these devices next to the usual unencrypted browsing done by mobile devices. This can be a significant privacy risk, even for seemingly innocuous data, such as light bulb states. Inferences such as presence in rooms or homes can be made just by inspecting the responses to state requests, e.g.\ in this case, the periodic requests by the Amazon Echo to the Hue Bridge. Other information such as who was the last to control the lights and when can also be extracted from these requests.

\subsubsection{Controlling Devices}

Some devices have rudimentary security, such as the Hue Bridge, which requires a physical button press to register a new device. However, when keys and credentials are transmitted over plain HTTP, as they are in the case of the Hue Bridge, any device in the home network can sniff these, making the bridge susceptible to replay attacks and the keys can be hijacked to make other API calls.

We tested this by extracting the iPhones API key from our network trace, and making API calls to query state and control lights from a separate laptop with a different MAC address. The requests remained valid even after several weeks during which the bridge was unplugged. This indicates that the key is persistent, has no short timeout, and not bound to any device by IP or MAC address.

While a malicious or compromised device, or an intruder in the network, might necessarily have any reason to control someone's lights, devices that emit more sensitive data, or control more critical systems, can be affected by similar vulnerabilities,

Attack surface is of course not limited to WiFi. It has been shown in the past that a Samsung Smart Things can be compromised through Zigbee \cite{fernandes2016security} for example. As we only captured network traffic, this is outside the scope of this paper.

\subsection{Global Statistics}

An initial analysis of our network trace showed that the largest source of data by IP in bytes is an Apple server (25.8\%), and correspondingly the top largest sink is the Apple TV (41.9\%). These two IPs account for 27.1\% and 43.9\% of our routers sources and destinations by bytes respectively. This is singlehandedly caused by a large software update, that equally had a significant effect on skewing global port statistics towards HTTP and HTTPS. By volume, just under half of traffic was as a result of this alone.

\begin{figure*}
	\centering
	\includegraphics[width=\columnwidth]{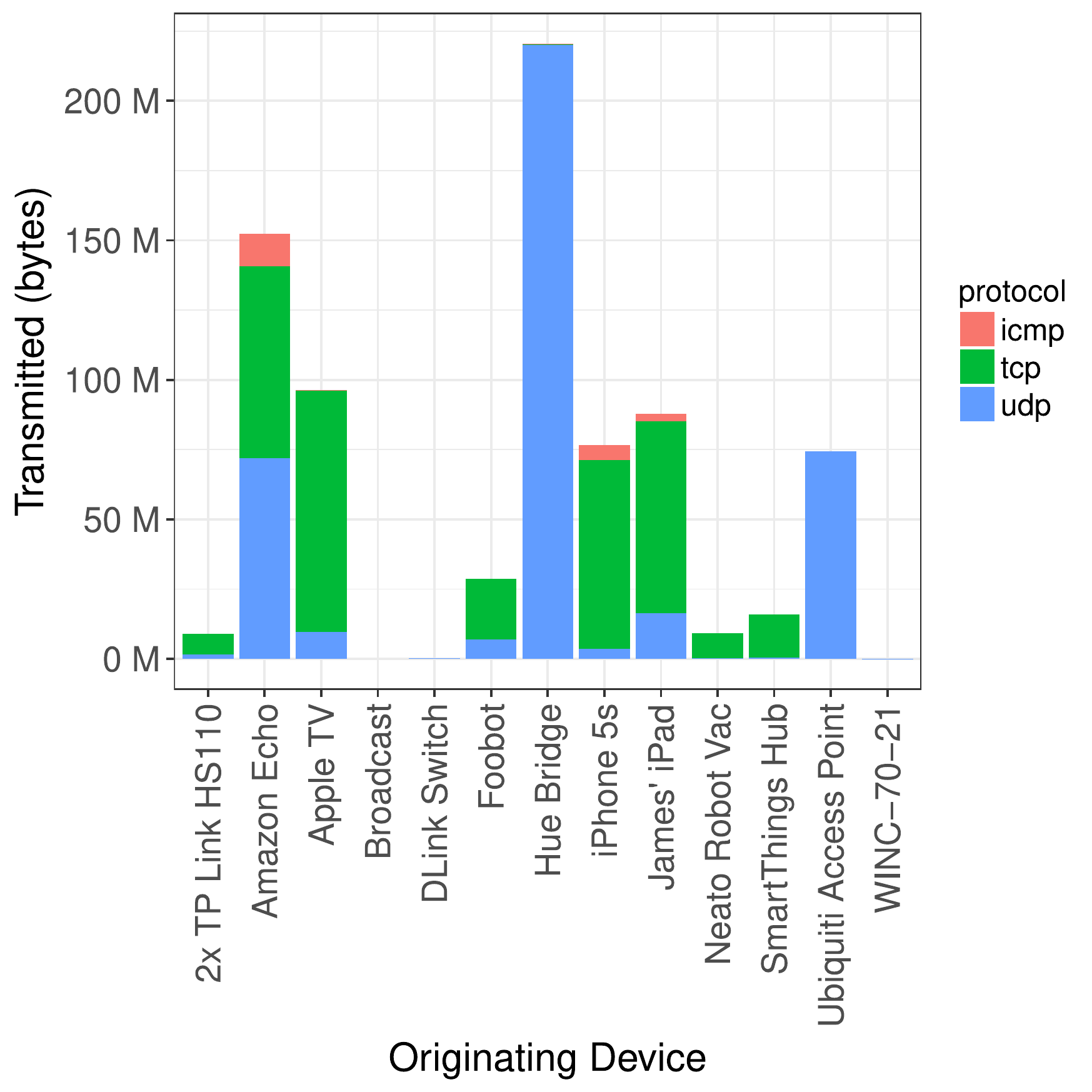}
	\includegraphics[width=\columnwidth]{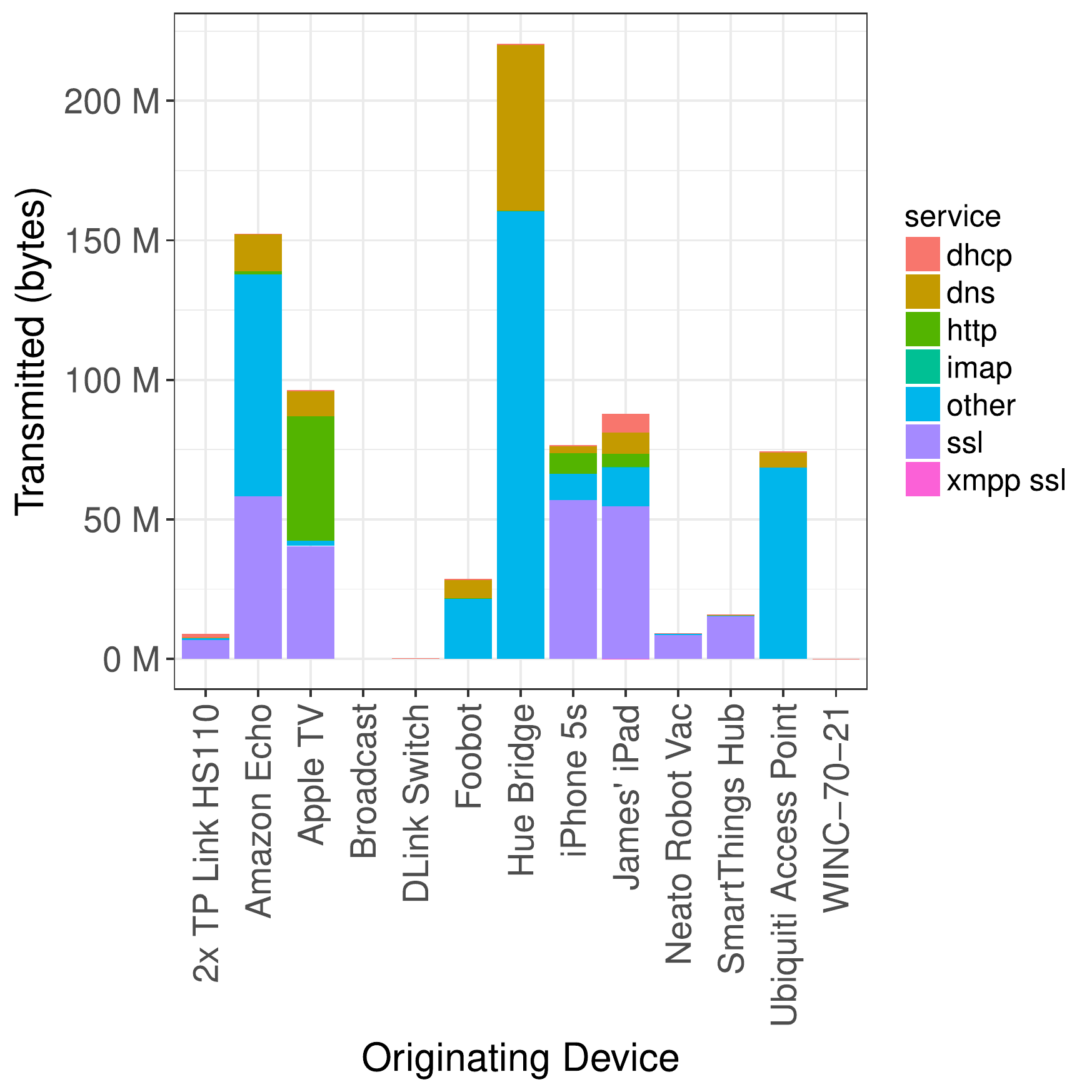}
	\caption{Bytes transmitted per device split by protocol and service}
	\label{f:traffic}
\end{figure*}

\begin{figure*}
	\centering
	\includegraphics[width=\columnwidth]{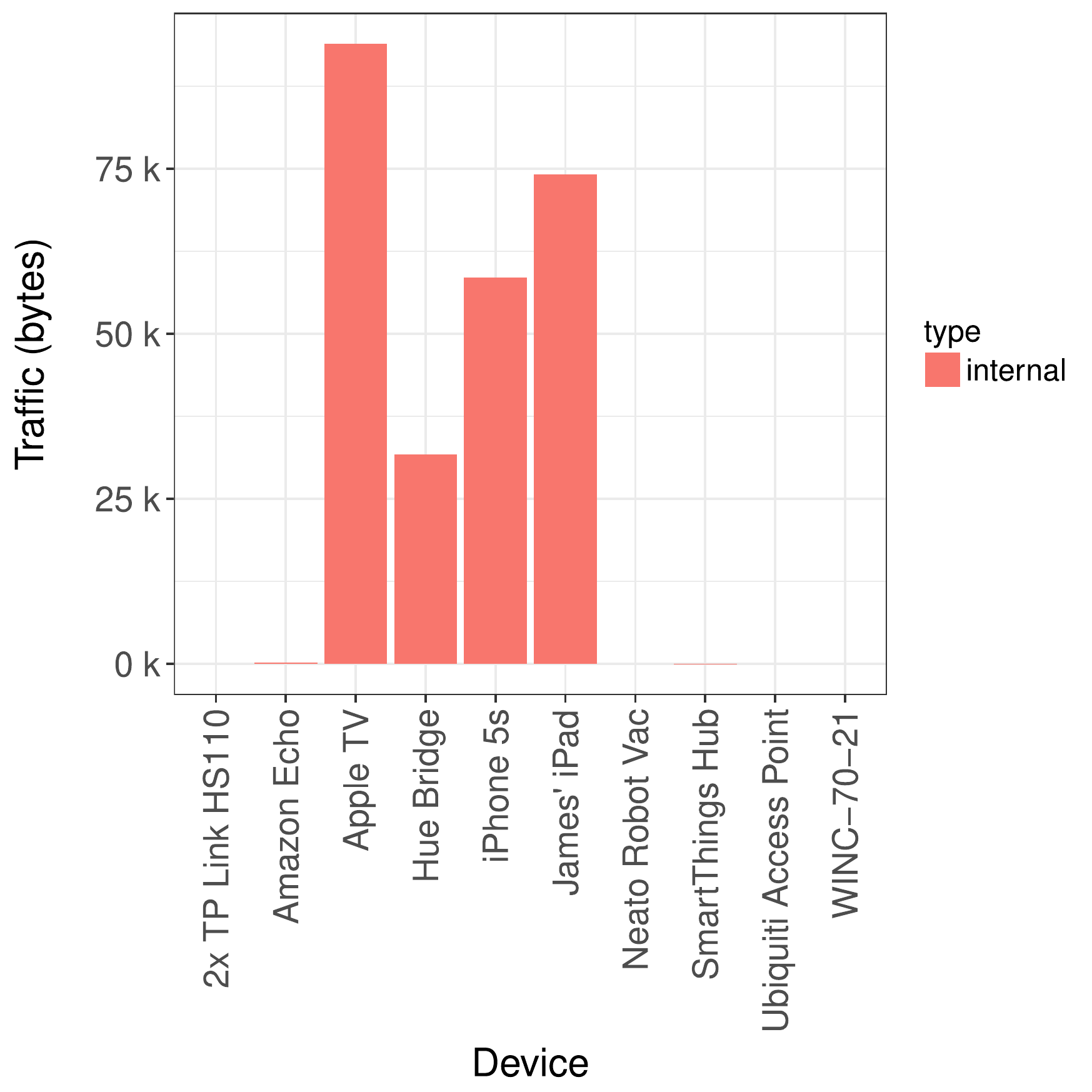}
	\includegraphics[width=\columnwidth]{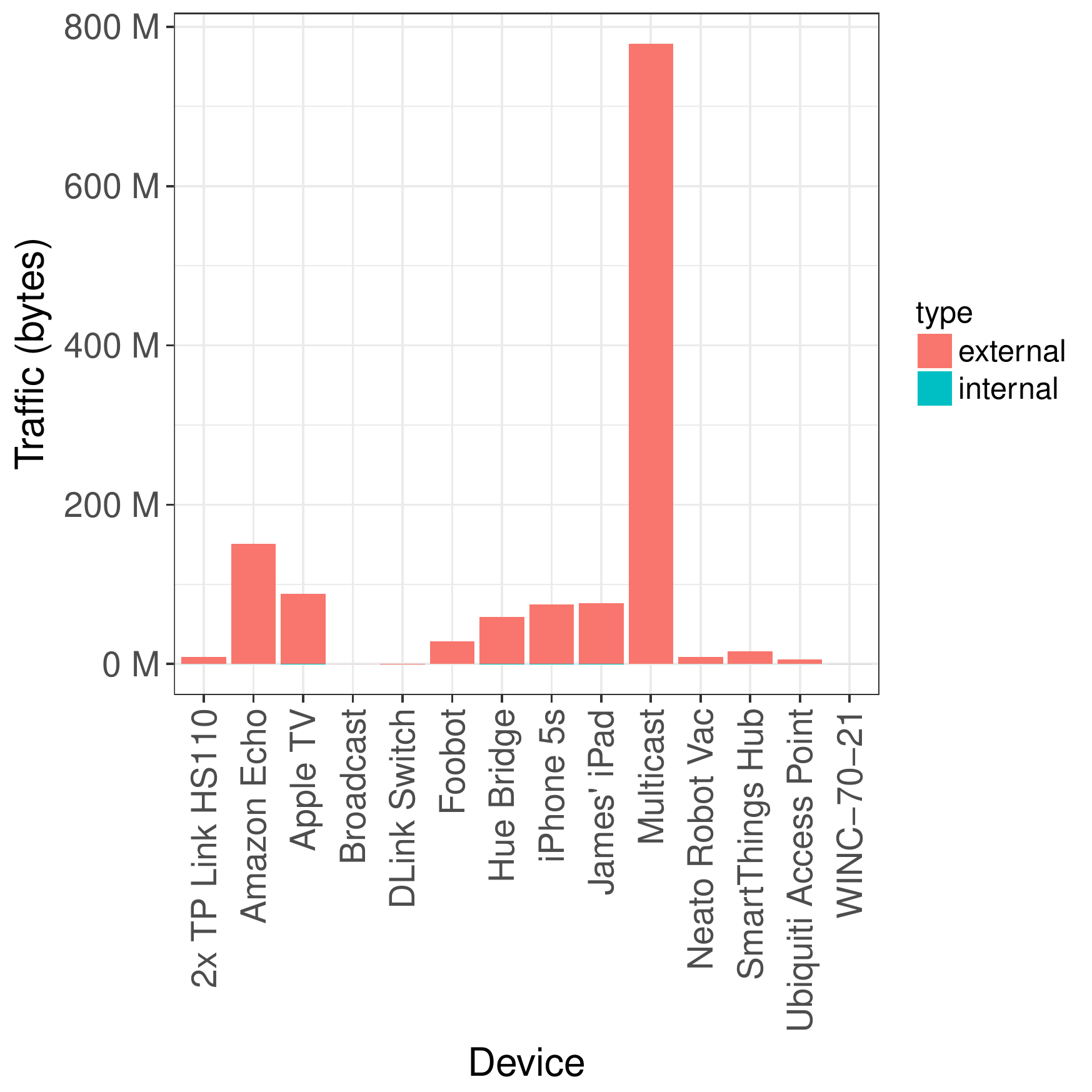}
	\caption{Internal vs external traffic; internal traffic separated for visibility}
	\label{f:internalexternal}
\end{figure*}

Figure~\ref{f:traffic} shows the bytes transmitted by device (taken from the IP total\_length header field) split by protocol (left) and service (right). Figure~\ref{f:internalexternal} shows the sum of payload bytes for each connection sent by originator and responder split by internal and external communication, the vast majority of which is external.

This information was extracted from Bro connection logs where each connection has an originator and a responder. Figure~\ref{f:traffic} only shows just the number of bytes an originator sent. Our TP Link MR6400 router accounts for virtually all bytes \emph{responded} at 4.02 GB in total, 1.99 of which is HTTP, 1.85 SSL, 0.09 DNS, a major part of which can be attributed to the Apple TV update. The router was left out of figure~\ref{f:traffic} as it dwarfs the remaining devices being responsible for the most originating bytes at 778.73 MB.

\newcommand*\circledwhite[1]{\tikz[baseline=(char.base)]{\node[shape=circle,draw,inner sep=1pt,scale=0.8] (char) {#1};}}

\begin{figure*}
	\centering
	\includegraphics[width=\linewidth]{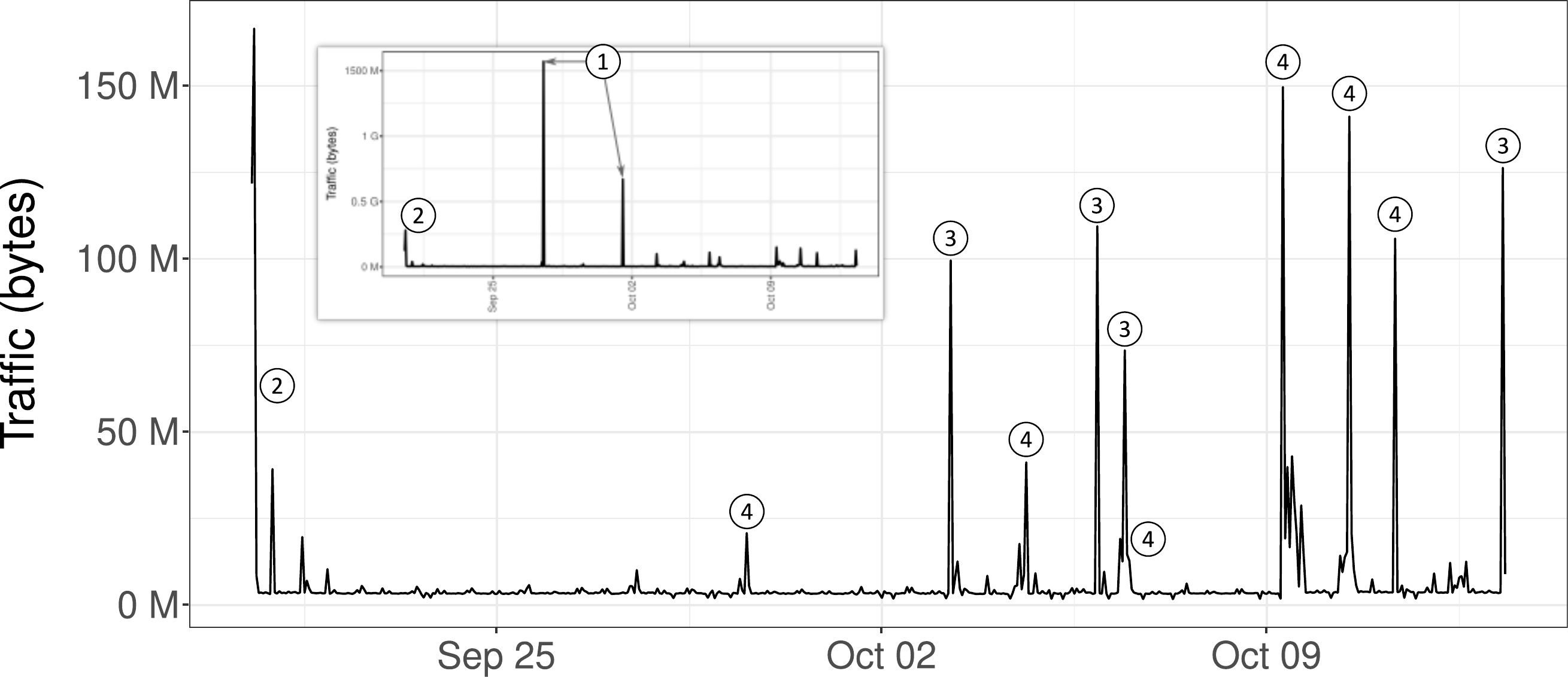}
	\caption{Hourly aggregates of frame lengths over time with spikes from Apple TV \protect\circledwhite{1}, iPhone \protect\circledwhite{3}, iPad \protect\circledwhite{4}, and mixed configuration \protect\circledwhite{2}}
	\label{f:packet-lens-hourly-combined}
\end{figure*}

This is especially visible in the inset plot in figure~\ref{f:packet-lens-hourly-combined} where the Apple TV update created large spikes at \circledwhite{1}. When these are filtered out, we are left with figure~\ref{f:packet-lens-hourly-combined} proper where most spikes are as a result of iPhone \circledwhite{3} and iPad \circledwhite{4} browsing. The initial spikes at the start of the trace \circledwhite{2} are caused by traffic during configuration and is discussed in the next section.

\begin{figure*}
	\centering
	\includegraphics[width=\linewidth]{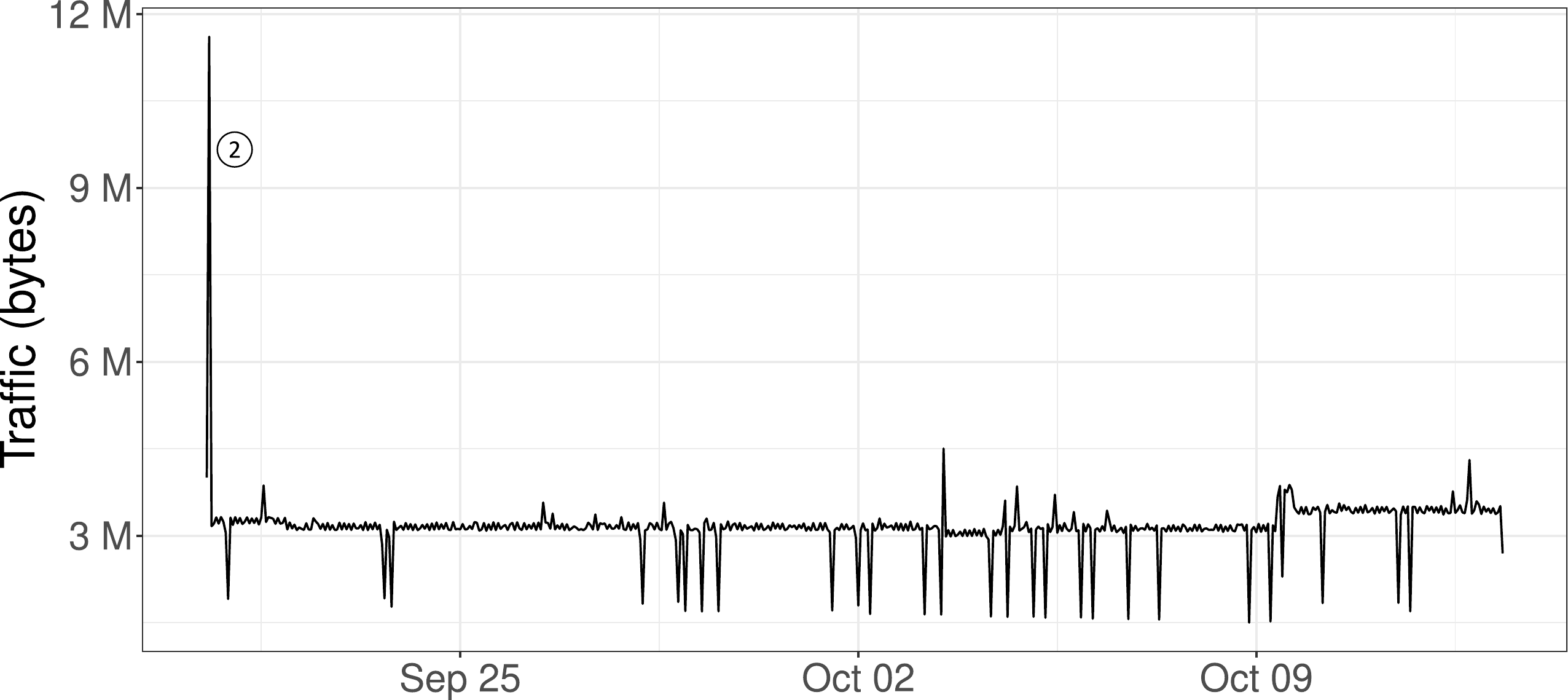}
	\caption{Hourly aggregates of frame lengths over time excluding Apple TV, iPhone, and iPad}
	\label{f:packet-lens-hourly-no-apple-tv-no-iphone-no-ipad}
\end{figure*}

\subsection{Statistics by Device}

\subsubsection*{Router}

Other than the Apple TV update and some MQTT pings, virtually all traffic originating at the router were from SSDP packets. We can tell this alone by destination, which was the SSDP multicast address. The Search Target headers contain device and service schemas that are typical of routers, for example, InternetGatewayDevice, WANDevice, WANCommonInterfaceConfig, WANConnectionDevice, WANIPConnection, WFADevice, WFAWLANConfig, and Layer3Forwarding. However no devices request UPnP descriptions from the URLs in the location headers.

\subsubsection*{Hue Bridge}

The Hue Bridge's behaviour is particularly interesting, as the vast majority of its traffic is consistent DNS and SSDP, after a brief burst in the beginning during configuration. This is primarily (787552 requests) to www.ecdinterface.philips.com, however it does not attempt to connect to this host.

The Hue Bridge transmits bursts of just over 2 KB of basic device SSDP traffic every 52 seconds, however, unlike the router, its UPnP description is queried by the Amazon Echo. The Echo gets standard metadata from this description, and at thrice the frequency, the Echo calls Hue API endpoints to get the lights' state, including on/off state, colour information, and other metadata. The Echo is the only device that makes API calls to the Hue bulbs, aside from the iPhone during configuration.

\subsubsection*{Apple TV}

Apple TV network activity can for the most part be attributed to the aforementioned automatic software update. Automatic updates can be disabled but are enabled by default. The unencrypted portion of this traffic are primarily spikes while requesting image and video thumbnails from Akamai Technology CDN servers.

These include film cover art and app icons, so viewing habits and installed apps can be inferred indirectly by examining the frequency certain thumbnails are downloaded. There exists therefore the potential for embarrassment on, for instance, cultural or religious grounds.

When looking at purely the number of connections by port, DNS has the highest at 4480 followed by SSL at 4434, and multicast DNS at 3580. From the DNS requests, we can see that the most frequent is a local HomeKit hostname, followed by Apple's Bonjour Sleep Proxy (common across any Apple device), and a number of Apple application and time servers (time-ios.g.aaplimg.com, time-ios.apple.com, itunes.apple.com, init.itunes.apple.com, xp.apple.com, play.itunes.apple.com, etc). There are no other discernible patterns.

\subsubsection*{Amazon Echo}

Aside from the traffic corresponding to Hue Bridge API and UPnP requests previously discussed, all HTTP traffic is encrypted. The Echo communicates with several Amazon servers. The DNS logs hint on the Echo's behaviour, with 4392 requests for device-metrics-us.amazon.com, 1210 for dcape-na.amazon.com, 538 for pindorama-eu.amazon.com, and 46 for softwareupdates.amazon.com, among several others. Interestingly, there were also 490 requests for www.meethue.com; the model URL listed in the Hue Bridge's UPnP description.

The vast majority of DNS requests were very unusual -- 69192 for www.example.com, 69120 for www.example.net, and 69086 for www.example.org. Other users have observed this, however we can only speculate what the point of these requests are, since the Echo's interaction with these hosts is limited to the TCP three-way handshake.

By number of connections, 29796 NTP connections follow DNS, with the most at 44973, followed by HTTP at 22090 connections. There were also 6778 ICMP connections made from port 8 on the Echo to port 0 on the router.

\subsubsection*{Neato Botvac}

In figure~\ref{f:traffic}, this device's total transmitted bytes are relatively small since it was turned off between 2017-09-22 01:20 and 2017-10-04 15:50. The vast majority of traffic is TCP; a baseline of SSL traffic with mostly TCP Keep-Alive packets. It communicates mainly with two AWS servers in the cloud. Browsing to these shows that these are Neato's RESTful API servers; ``Nucleo'' servers with version 1.11.1 of a ``Neato Robotics Message Bus.''. Other than that, there is minor NTP traffic.

\subsubsection*{iPhone}

Besides using device apps for configuration and setup at the start, subsequent iPhone traffic is characteristic of normal browsing behavior. There are random spikes of HTTP/SSL traffic to email, social media, and news websites, as well as the common Apple servers (akamaitechnologies.com CDN, etc). Some requests are to Amazon, attributable to Alexa.

\subsubsection*{Macbook Pro}

This was the first device that had Dropbox traffic. It was only really active during the first two hours of capture for configuration, and had barely any traffic at all after that, except short periods of primarily SSL, some HTTP traffic. Coupled with DNS requests to e.g.\ Facebook and Google servers, this is indicative of light browsing.

Some of the HTTP traffic was requesting \texttt{generate\_204} characteristic of network portal detection for logging in to WiFi, otherwise JSON headline dumps from news websites and some images.

Here too we have the usual Apple traffic, including sleep proxy and HomeKit, as well as Hue Bulbs interaction. The Macbook also had some encrypted communication to a second Netatmo server (b90.netatmo.net as opposed to b91.netatmo.net).

\section{Conclusions and Ongoing Work}\label{s:conclusions}

In this work, we analyse network traces from a testbed of common IoT devices, and describe general methods for fingerprinting their behaviour. We then use the information and insights derived from this data to assess where these privacy and security risks manifest themselves, as well as how device behaviour affects bandwidth and power consumption.

We published scripts to simplify doing this form of network trace analysis and identified several areas of contention when it comes to privacy such as Philips Hue Bridge security holes and simple circumvention of MAC address randomization. The main takeaway from this work is to begin to provide an awareness on the behavior of common home IoT devices to mitigate privacy risks.

Our initial analysis was on a simple case where all IoT devices are in their idle state. We plan to repeat this analysis for several additional scenarios and under different conditions.

Through observing behavior when users directly interact with these devices at home and remotely (e.g.\ through an app) we seek to ultimately draw inferences from the traffic, such as discerning human activity and evaluating the privacy risk of being able to do so. Similarly, we plan to lay the groundwork for identifying important considerations when building integrated IoT/home hub systems from a network security perspective.

\section{Acknowledgements}
This work is funded in part by the grant EP/M001636/.

\bibliographystyle{IEEEtran}
\bibliography{IEEEabrv,tma18}

\end{document}